\documentclass{article}
\usepackage{graphicx} 

\usepackage[numbers,sort&compress]{natbib}

\usepackage{multicol}
\usepackage[left=2.5cm, right=4cm]{geometry} 
\usepackage{placeins}

\usepackage{xcolor}
\usepackage{soul}
\usepackage{authblk}
\usepackage{etoolbox}
\usepackage[dvipsnames,svgnames,x11names]{xcolor}

\title{DeepConf: Machine Learning Conformer Reconstruction of Biomolecules from Scanning Tunneling Microscopy Images}
\author[1]{Tim J. Seifert}
\author[2]{Dhaneesh Kumar}
\author[1,3]{Markus Etzkorn}
\author[4,5]{Stephan Rauschenbach}
\author[2]{Klaus Kern}
\author[2]{Kelvin Anggara}
\author[1,3]{Uta Schlickum}

\affil[1]{Institute of Applied Physics, Technical University Braunschweig, Braunschweig, Germany}
\affil[2]{Max Planck Institute for Solid State Research, Stuttgart, Germany}
\affil[3]{Laboratory for Emerging Nanometrology LENA, Braunschweig, Germany}
\affil[4]{Department of Chemistry, University of Oxford, Oxford, UK}
\affil[5]{Kavli Institute for Nanoscience Discovery, University of Oxford, Oxford, UK}
\date{}
\usepackage{siunitx}
\usepackage{subcaption}

\DeclareSIUnit\pixel{px}

\usepackage[acronym]{glossaries}
\newacronym{AFM}{AFM}{Atomic Force Microscopy}
\newacronym{ESIBD}{ESIBD}{Electrospray Ion Beam Deposition}
\newacronym{STM}{STM}{Scanning Tunneling Microscopy}
\newacronym{ML}{ML}{Machine Learning}
\newacronym{SPM}{SPM}{Scanning Probe Microscopy}
\newacronym{STEM}{STEM}{Scanning Transmission Electron Microscopy}
\newacronym{DFT}{DFT}{Density Functional Theory}
\newacronym{HR-AFM}{HR-AFM}{High-Resolution Atomic Force Microscopy}
\newacronym{UFF}{UFF}{Universal Force Field}
\newacronym{CNN}{CNN}{Convolutional Neural Network}
\newacronym{AI}{AI}{Artificial Intelligence}
\newacronym{UHV}{UHV}{ultrahigh vacuum}
\newacronym{MDS}{MDS}{multi-dimensional scaling}
\newacronym{STRESS}{STRESS}{Standardized Residual Sum of Squares}
\newacronym{UMAP}{UMAP}{Uniform Manifold Approximation and Projection}
\newacronym{kNN}{kNN}{k-nearest neighbors}
\newacronym{MD}{MD}{Molecular Dynamics}
\newacronym{cryo-EM}{cryo-EM}{cryo-electron microscopy}

\DeclareSIUnit\angstrom{\text{Å}}

\begin{document}

\twocolumn[
\maketitle
\begin{abstract}

    Improving the detailed understanding of the underlying properties and functions of biomolecules has recently attracted growing interest, enabled by the possibility of real-space imaging of single, intact macromolecules using \gls{STM} in combination with electrospray ion beam deposition and soft landing. This combination provides key insights into biomolecular behavior, but it also imposes stringent requirements on rapid and reliable data analysis. A major limiting factor for applying machine learning to STM images is often the scarcity of training data, caused by the long acquisition times required for both experimental imaging and high-accuracy simulations. Here, we propose a framework for the rapid generation of three-dimensional structures of glycans, peptides, and glycopeptides and their corresponding STM-like image simulations, based on state-of-the-art, machine-learning–accelerated \gls{DFT}. We generate datasets for the polypeptide bradykinin and for a representative glycan molecule, and we train a conformer estimation model to predict a molecule’s three-dimensional structure from an STM image. On synthetic data, our approach achieves high accuracy, with median atomic deviations below \SI{2}{\angstrom} for peptides and below \SI{4}{\angstrom} for glycans. Application to experimental data predominantly yields a precise, reliable, and visually convincing determination of the local positions of molecular subunits. The application to experimental data represents an important milestone towards a fully automated structural search pipeline for complex, biologically relevant systems imaged with STM.
\end{abstract}
\glsresetall
\vspace{1em} 
]

The investigation of biomolecules, like peptides and glycans, is essential for understanding fundamental processes across all forms of life\cite{unwin_molecular_1975, wu_imaging_2020}. Peptides play key roles in living organisms as antibiotics \cite{gao_defensins_2021, hancock_peptide_1999} and through their structural\cite{cheng_beta-peptides_2001} and signaling functions\cite{silberbach_natriuretic_2001, matilla_catalogue_2022}. Glycans are primarily found covalently attached to lipids forming glycolipids, or to proteins as glycoconjugates, where they govern cellular growth\cite{shivatare_glycoconjugates_2022, fukuda_cell_1984}, function\cite{shivatare_glycoconjugates_2022, yanagisawa_expression_2007}, and energy storage\cite{shivatare_glycoconjugates_2022, kang_carbohydrate_2015}.
For a long time, biomolecules could only be studied through en\-sem\-ble-averaged measurements, especially \gls{cryo-EM}. Cryo-EM is able to resolve single proteins with remarkable resolutions, but requires averaging images of many individual and identical molecules\cite{bai_how_2015, zhong_cryodrgn_2021}. Thus these ensemble averaging methods cannot be extended to flexible molecules appearing in different conformations. Hence, for many biologically relevant molecules, imaging individual molecules at the submolecular level is indispensable for obtaining crucial structural insights due to their internal flexibility and structural heterogeneity, allowing them to adopt a wide variety of three-dimensional conformations\cite{huber_conformational_1979, mierke_peptide_1994, callaway_revolution_2015}. Recently, direct observation of single glycoconjugate mo\-le\-cu\-les has become feasible by combining \gls{STM} with \gls{ESIBD}\cite{rauschenbach_mass_2016}, opening new perspectives for the study of biomolecules\cite{wu_imaging_2020, anggara_direct_2023}. This conformational flexibility is directly observed in \gls{STM}, where markedly different geometries can be found for identical molecules. However, the interpretation and structural decomposition of such images is often non-trivial due to conformational ambiguity and inherent imaging limi\-ta\-tions\cite{wu_imaging_2020, anggara_direct_2023, struwe_high-resolution_2024}. Consequently, manual analysis requires substantial expertise, yields user-dependent results, and is extremely time-con\-su\-ming\cite{choudhary_computational_2021, alldritt_automated_2020, maksymovych_methanethiolate_2006}.

These challenges can be addressed by employing \gls{ML} for automated structural analysis. In recent years, \gls{ML} has been increasingly integrated into surface science, facilitating the use and data analysis of \gls{SPM} and \gls{STEM}. Major applications include automated microscope oper\-ation\cite{sotres_enabling_2021, kalinin_automated_2021, krull_artificial-intelligence-driven_2020}, image enhancement\cite{moller_super-resolution_2023, liu_general_2019}, and image analysis\cite{seifert_chirality_2024, li_machine_2021, kurki_automated_2024, carracedo-cosme_molecular_2024, alldritt_automated_2020}. 

The latter can be divided into two main directions. First, \gls{ML} has been used to detect and classify features such as individual molecules\cite{li_machine_2021} and molecular clusters\cite{seifert_chirality_2024} in large-scale \gls{SPM} images. Second, several pioneering efforts have been made toward automatic structure determination from single-molecule images\cite{kurki_automated_2024, carracedo-cosme_molecular_2024, alldritt_automated_2020}. 
Using three-dimensional image stacks of \gls{DFT}-simulated \gls{AFM} images, high resolution molecular structures have been reconstructed\cite{alldritt_automated_2020}. This approach is extended by applying conditional generative adversarial networks to AFM images of small and planar molecules\cite{carracedo-cosme_molecular_2024}, and further  generalized to predict atomic positions in bond-re\-sol\-ved STM images using a \gls{CNN} architecture\cite{kurki_automated_2024}.

In all these approaches, functionalized tips enable the direct resolution of individual bonds between atoms, providing detailed submolecular information. However, when the molecules of interest are no longer planar, as is typically the case for biomolecules, they adopt complex three-dimensional conformations. In that case, or if non-functionalized \gls{STM} tips are used, they so far cannot be imaged with comparable resolution; thus, very limited intramolecular detail is visible for these molecular structures\cite{wu_imaging_2020}.

A critical bottleneck common to all \gls{ML} applications for \gls{AFM} and \gls{STM} images is the inherently slow acquisition speed of \gls{SPM}. Robust \gls{ML} models require large, diverse datasets for training, which cannot be generated within a reasonable time frame by \gls{SPM} measurements and manual labeling alone. 

In previous work\cite{seifert_chirality_2024}, we relied exclusively on fully computer-generated synthetic data to train an object detection model. This model successfully detected and classified chiral unit cells in real \gls{STM} images of a self-assembled molecular network, demonstrating that training on a diverse synthetic dataset can yield highly robust performance on experimental data\cite{seifert_chirality_2024}, clearly outperforming approaches based solely on the augmentation of real images\cite{li_machine_2021}.

Considering the investigation of large, biologically relevant macromolecules, simple rigid molecular models are no longer sufficient and require a fundamentally different strategy for automatic structure discovery. 
We demonstrate that by adapting both the data generation procedure and the machine learning (ML) setup, these limitations can be overcome. This enables the automatic estimation of three-dimensional complex molecular conformations of biomolecules in \gls{STM} images. 

\begin{figure*}[h]
    \centering
    \includegraphics[width=0.8\textwidth]{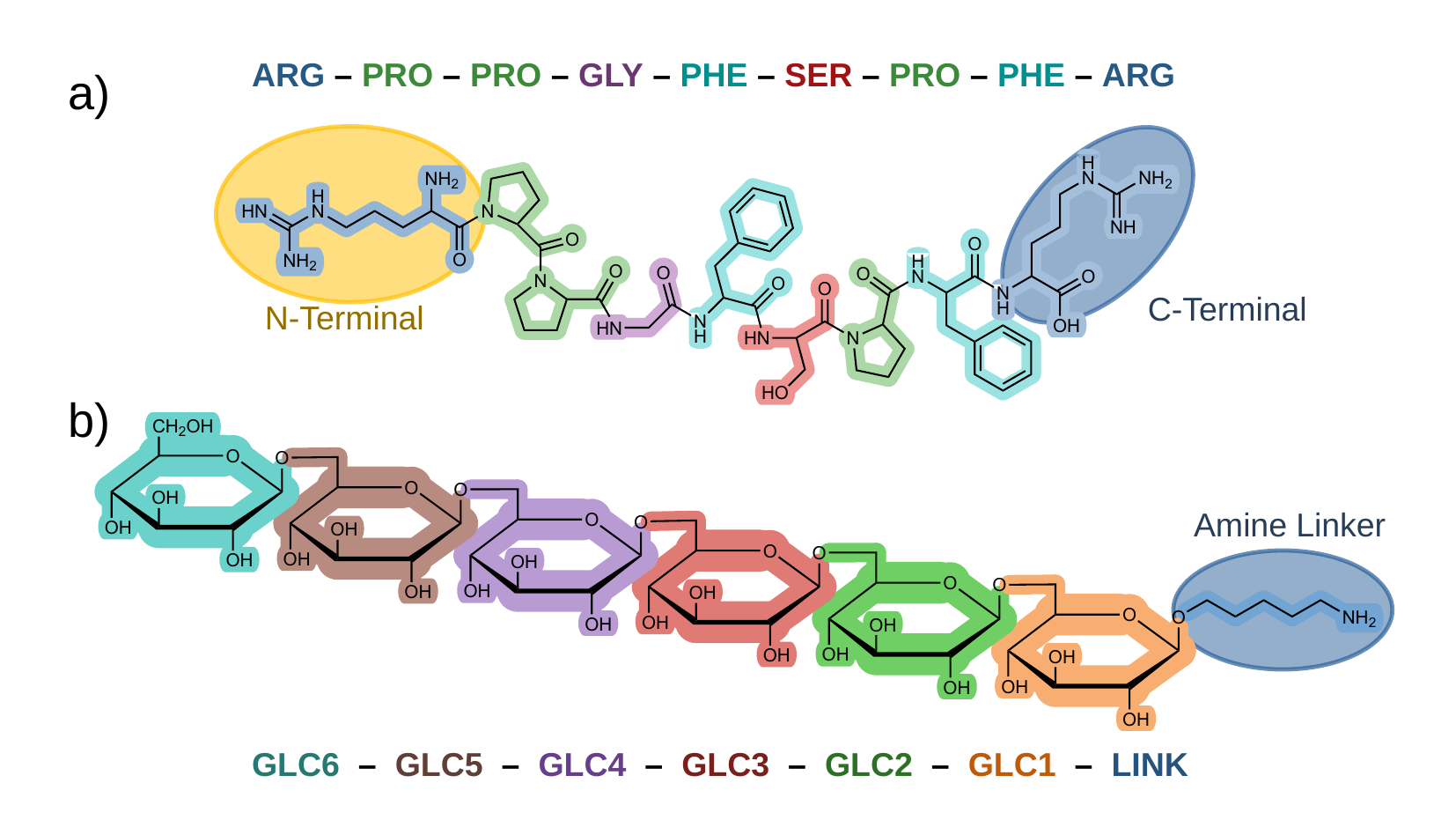}
    \caption{a) Bradykinin molecule used as an exemplary peptide. The chain is color coded according to the individual amino acids. Identical amino acids have the same color. The N- and C-termini are marked with an orange and a blue overlay respectively. b) Amine terminated glucose hexamer used as an exemplary glycan. The chain is color coded for each monosaccharide and the amine linker.  The linker is marked with a blue overlay.}
    \label{fig: Structures}
\end{figure*}

We present results for two exemplary mo\-le\-cu\-les: the polypeptide bradykinin, which consists of nine amino acids (ARG–PRO–PRO–\-GLY–\-PHE–\-SER–\-PRO–\-PHE–\-ARG) \cite{rauschenbach_two-dimensional_2017}, and as an exemplary glycan a beta-(1,6) glucose hexamer which is terminated with a penta-1-amine unit (amine linker)\cite{wu_imaging_2020}, see figure \ref{fig: Structures} for reference.

On synthetic data, we find excellent agreement between the predicted overall molecular structure and the ground-truth conformation used to generate the images. Going beyond overall structure estimation, we quantify the model’s ability to predict the direct coordinates of all atoms in the synthetic images, obtaining median distances between true and predicted atomic positions below \SI{2}{\angstrom} and \SI{4}{\angstrom} for the exemplary peptide and glycans, respectively. On experimental \gls{STM} data, the predicted overall molecular structures also show strong consistency with the measurements. For the polypeptide molecules, this, in turn, enables automatic structural classification with high accuracy, as demonstrated on both synthetic and real datasets. We conclude that our method goes far beyond current approaches to automatic structure discovery by enabling the analysis of complex, non-planar macromolecules. This represents a major step towards fully automated molecular structure investigation in \gls{STM}.

\section*{Results}

\subsection*{Training Data Generation}
\begin{figure}[tbph]
    \centering
    \includegraphics[width=0.45\textwidth]{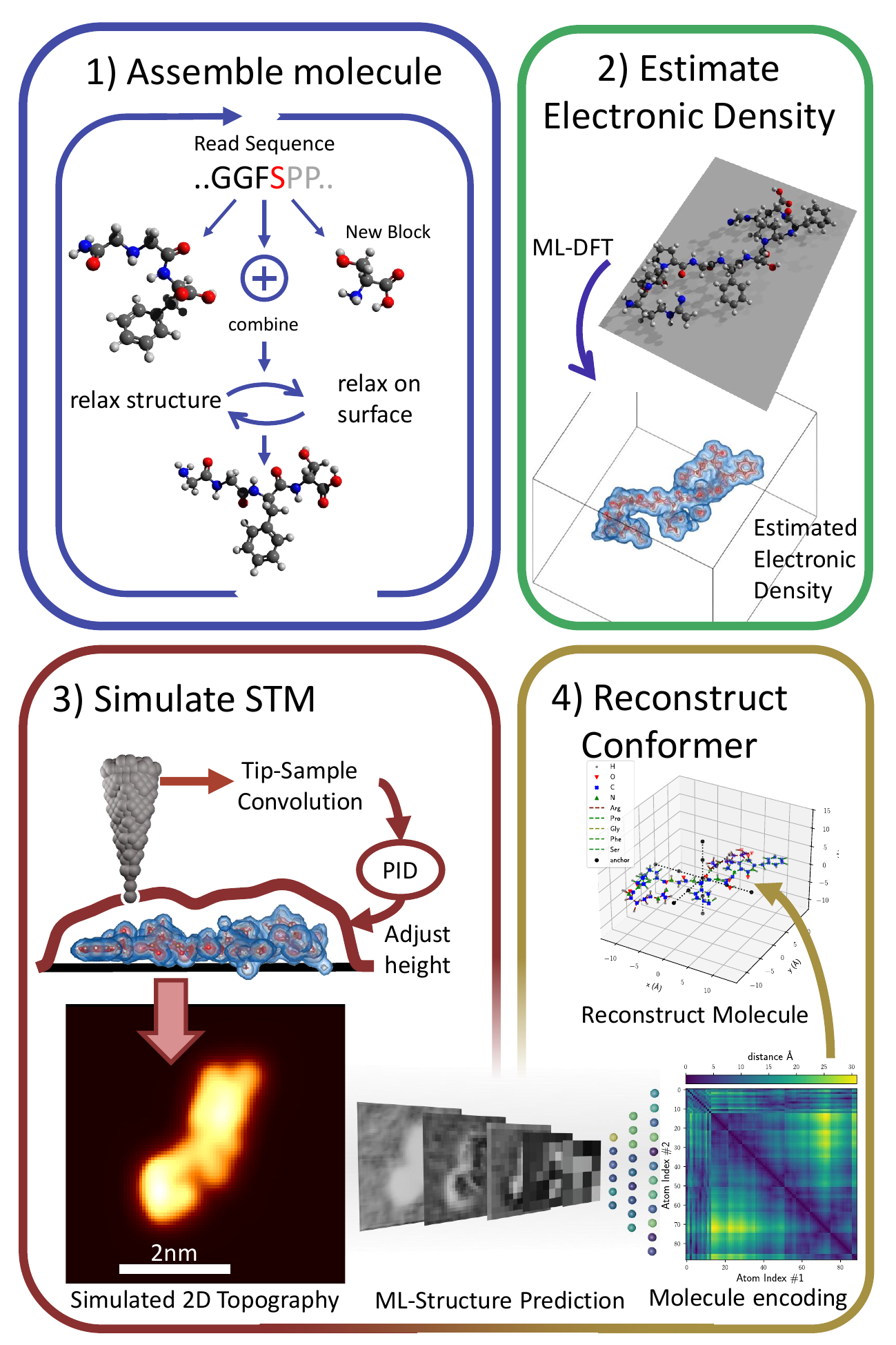}
    \caption{Schematic Workflow of the image generation process. First, (1) a random molecular geometry is created by parsing the molecular sequence, attaching one amino acid or monosaccharide at a time, followed by structural and surface relaxation steps. Secondly, (2) the electronic density is estimated using \gls{ML} emulated \gls{DFT}\cite{del_rio_deep_2023}. With this (3),  a \gls{STM} image is extracted, created by convolving a random tip geometry with the electronic density. This image can be used to train the \gls{ML} model to predict a molecular encoding, which can be reconstructed to retrieve the 3D molecular structure (4).}
    \label{fig: DataGenScheme}
\end{figure}

The collection of high-quality training data is essential for the success of any \gls{ML} approach. Figure \ref{fig: DataGenScheme} provides a schematic overview of the synthetic data generation process used to simulate \gls{STM} images and the \gls{ML}-workflow. Our data generation pipeline robustly produces plausible looking molecular conformations and simulates realistic \gls{STM}-like images for \gls{ML} training in a unified workflow. This process is implemented for peptides, glycans, and glycopeptides with example structures shown in the \textit{Supplementary Information} \ref{SI_Data}. The data generation procedure consists of three main modules: (i) generating molecular conformations, (ii) estimating their electronic densities (iii) simulating \gls{STM}-like images.

\subsubsection*{i) Generating molecular conformations}
The first step in the data-generation procedure is to generate a large number of distinct molecular conformations that can be used for training, given a specific peptide, glycan, or glycopeptide structure. The resulting molecular geometries must be both realistic and highly diverse in order to represent the range of structures that may be observed in experimental images. Data generation is implemented iteratively: the molecular structure is constructed block by block. Individual blocks correspond either to single amino acids or single glycan units. Each new block is appended to the end of the growing chain, or, in the case of glycans or glycopeptides, it can branch off from the main chain. At initialization, the structure of the newly attached amino acid or monosaccharide is defined by its predefined standard geometry. The connection to the existing chain is made in an randomly chosen orientation, allowing for any plausible bonding angle, thus leading to the generation of highly diverse conformations. 

The newly extended molecule is initially strai\-ned at the non-native bonding angles; therefore, the structure must be relaxed to yield physically meaningful conformations. In contrast to other work\cite{guo_molecular_2022}, we do not explicitly search for the conformation with the lowest possible energy as an isolated molecule. Due to electronic interactions with the surface, alignments that differ from gas-phase optimized conformations are commonly observed in STM experiments, particularly because adsorbed molecules tend to strive for stronger contact with the surface, hence producing more two-dimensional conformations compared to those in their native environment\cite{mahadevi_cation-pi_2013}. Explicitly calculating the electronic interactions between the molecule and the surface would be computationally prohibitive. Instead, we employ an iterative molecular relaxation scheme based on the \gls{UFF} to optimize the gas-phase conformation, combined with a simplified description of surface relaxation modeled via a modified Lennard–Jones potential. This procedure yields semi-optimized geometries that approximate molecular adsorption on the surface. 

\subsubsection*{ii) Estimating electronic density}
For each generated conformer, the electronic density is predicted using a \glsentrylong{ML} surrogate of \glsentrylong{DFT} \cite{del_rio_deep_2023}. \textit{Del Rio et al.} trained a \gls{ML} model to approximate the results of full \gls{DFT} calculations for a broad range of organic molecules, polymer chains and crystals containing the same atomic components as our molecules. Thus on the very localized scale taken into account in the \gls{ML}-\gls{DFT}, their training structures are similar to the biomolecules assessed in this study. Their implementation already achieves a substantial speed-up relative to conventional electronic-structure methods. With additional technical optimizations, we further reduced the computation time to as little as 10 seconds per structure for molecules containing up to 150 atoms on a single \textit{NVidia RTX 3080} GPU (see \textit{Methods} for details).

\subsubsection*{iii) Simulating \gls{STM}-like images}
The predicted electronic structure can be used to compute multiple \gls{STM}-like images under different imaging conditions. To simulate \gls{STM} images, we use a simplified model to calculate the overlap between the molecular electronic density and a modeled tip. From this overlap, a topographic image is derived in a manner analogous to constant-current \gls{STM} imaging. 

The \gls{STM} tip is modeled as a stack of rotated, two-dimensional Gaussian functions with increasing radii. By varying the tilt, eccentricity, and effective tip radius, we enhance the diversity of the simulated dataset. 
The convolution of this tip model with the molecular electronic density, together with a representation of the metal surface as a simple exponential charge decay, is passed to a PID controller to simulate constant-current \gls{STM} operation at a per-image randomly chosen setpoint. By randomly sampling the tip parameters and the setpoint, multiple distinct images can be generated from  a single \gls{DFT} result, further increasing the volume and diversity of training data (see \textit{Methods} for details).

\subsubsection*{Generated training data}
We demonstrate our training-data generation and \gls{ML} workflow on two example systems: the poly\-peptide bradykinin and an amine terminated glucose hexamer.
As discussed by \textit{Rauschenbach et al.} \cite{rauschenbach_two-dimensional_2017}, bradykinin deposited by \gls{ESIBD} on a Cu(100) surface and imaged by \gls{STM} predominantly adopts three distinct surface conformations. 
We can incorporate this knowledge into our training data by adjusting the distribution of angles between adjacent amino acids. Apart from full random sampling across the entire sensible angle range, we additionally include conformations whose structural angles are sampled to be similar to those of the observed conformations. We denote the uniformly sampled molecules as class \textit{X} and those drawn from the three specialized angular distributions as classes \textit{A}, \textit{AB}, and \textit{B}. We provide a comprehensive set of generated \gls{STM}-like images in the \textit{Supplementary Information \ref{SI_Data}}. 

First, it can be concluded that the proposed conformer generation method indeed produces a broad range of conformations, yielding distinct shapes and adsorption geometries on the surface for both peptides and glycans. Looking at the angular distributions between individual amino acids, peptides from class \textit{X} exhibit very high structural diversity apparent from the broad angular distributions for each connection. The specialized classes more closely resemble the experimentally observed conformations but are still designed to cover a large range of possible arrangements. They differ in a few key connections, in which the bonding angle distribution is less broad and offset compared to other classes (see \textit{Supplementary Information \ref{SI_TrainDataStat}}). Secondly, the synthetically generated peptide conformers generally adsorb more or less in a flat orientation on the surface, underscoring the effectiveness of the surface-relaxation procedure. The atomic height distribution features a narrow peak around the minimum of the potential, with a total of \SI{90}{\percent} of atoms within a height range of \SI{3.1}{\angstrom}. In contrast, the glycans tend to adopt more three-dimensional arrangements, in agreement with experimental observations. In this case, the height distribution is significantly wider, locating \SI{90}{\percent} of atoms in a height range of \SI{6.9}{\angstrom} (see \textit{Supplementary Information} \ref{SI_TrainDataStat}). Third, the overall appearance of the simulated \gls{STM} images varies significantly from image to image, covering the range of appearances observed in experimental data. Together, the conformer generation strategy and the sampling of imaging parameters produce images that are both highly diverse and physically realistic, making them well-suited for constructing a robust training dataset. To achieve a robust \gls{ML} model that can also handle noisy real data, we alter the synthetic images with random rotations, flips, \gls{STM}-typical shot noise, line-wise noise, as well as background variation, thereby increasing the variance in the dataset and reducing the risk of overfitting.

\subsection*{Structure discovery workflow}
We employ a ResNet-based\cite{he_deep_2015} regression model to estimate molecular conformations from \gls{STM} images of single molecules. The prediction model takes an STM image as input and outputs a distance vector, from which the molecular shape and position can be reconstructed. For further details on the model architecture and training procedure, we refer to the \textit{Methods} section.  

\subsection*{Peptides}

To train the conformer prediction model, we generated a total of 5,000 synthetic peptide structures. For each, four different \gls
{STM}-like images were created with the same molecular structure, but different image generation parameters, mainly tip properties, yielding 20,000 training images in total.

\begin{figure*}
     \centering
    \includegraphics[width=\linewidth]{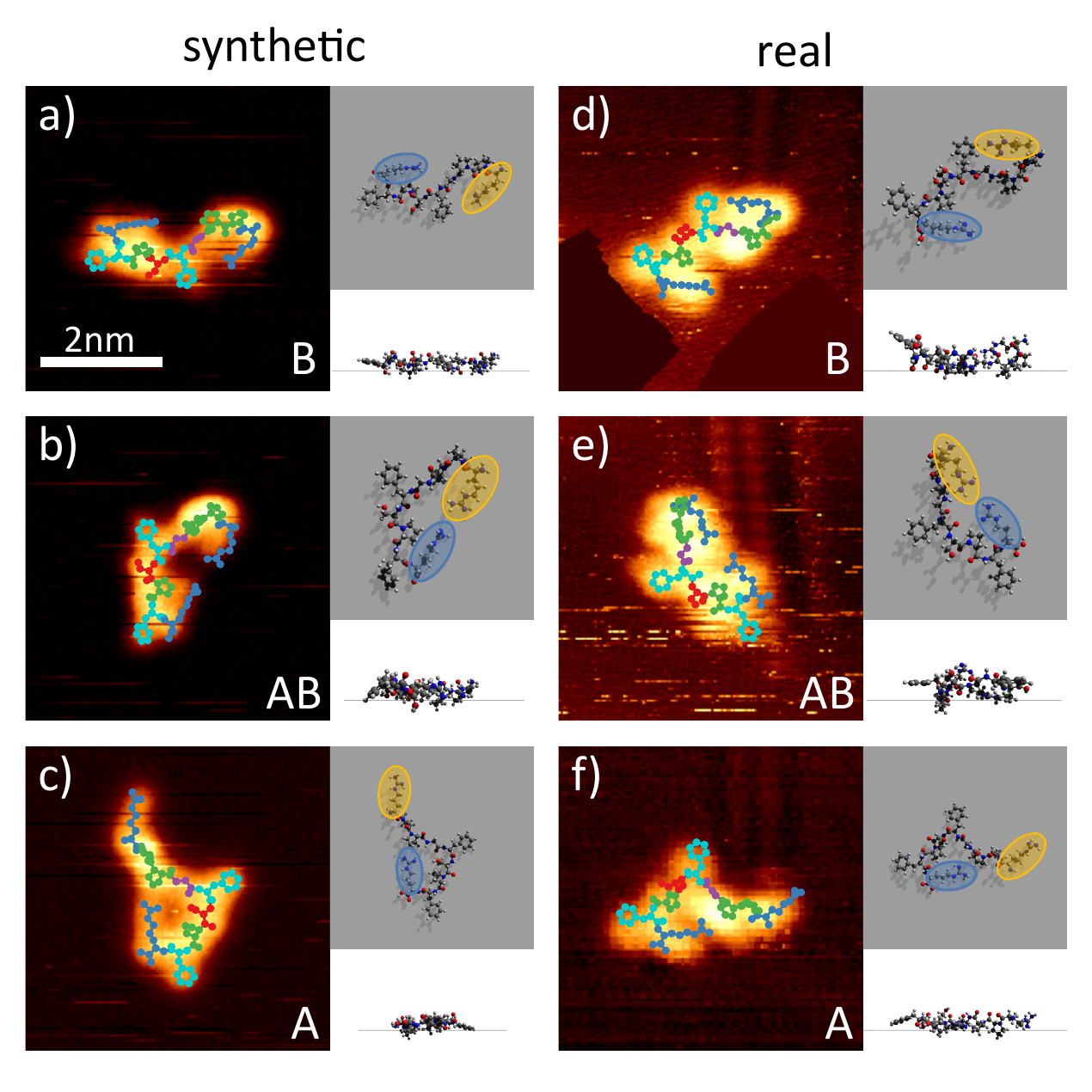}
     \caption{Results of conformer prediction model on synthetic and real peptide images. The overlay shows the flat two-dimensional projection of the predicted atomic positions. Additionally, top and side perspectives are shown. The amino acids are color coded and the arginine residues highlighted as in figure \ref{fig: Structures}. All images have a size of \SI{5}{\nano \meter} $\times$ \SI{5}{\nano \meter}.}
     \label{fig: PeptideResults}
\end{figure*}

\subsubsection*{Results on synthetic peptides}
The conformer prediction model was trained on the synthetic peptide dataset and then evaluated on 2000 images of 500 held-out synthetic test structures. The resulting predictions are shown in figure \ref{fig: PeptideResults} a)-c). In each example, the predicted conformer is visualized as a set of points corresponding to the non-hydrogen atoms, projected onto the image plane and color-coded by amino-acid type. The synthetic \gls{STM}-like images in this evaluation have been subjected to augmentations and noise, as was also done for the training data. For all examples shown, the predicted conformations closely reproduce the overall molecular shape visible in the \gls{STM} images, demonstrating the effectiveness of our approach for identifying the variety of conformations. Having established that the global molecular shapes can be reliably recovered, we next assess how accurately individual atom positions are predicted. Because each synthetic \gls{STM} image is generated from a known molecular geometry, the exact position of every atom in the image is available as ground truth. We compare these true positions with the locations predicted by the \gls{ML} model and quantify the deviation for each atom $i$ by computing the Euclidean distance between its predicted $\hat{x}_{m,i}$ and true position $x_{m,i}$. For each molecule, we then calculate the average distance of these distances across all $N_{\mathrm{at}}$ atoms. 
To increase the robustness of our metric against a few outliers, we determine the median distance $d$ over all synthetic peptide structures $m$ in the test set.
\[
d \;=\; \operatorname{median}_{m=1,\dots,M}\left(\frac{1}{N_{\mathrm{at}}}\sum_{i=1}^{N_{\mathrm{at}}}\left\lVert x_{m,i}-\hat{x}_{m,i}\right\rVert_2\right).
\]
We obtain a median atomic localization error of \protect\SI{1.5}{\angstrom}, while \protect\SI{20}{\percent} of images have an error below \protect\SI{1.2}{\angstrom} and \protect\SI{80}{\percent} an error below \protect\SI{2.5}{\angstrom}.
This demonstrates that our method not only captures the overall molecular shape but also localizes individual atoms in the \gls{STM} images with remarkably high accuracy.

\subsubsection*{Results on real peptides}
We now apply the model, which has been trained exclusively on synthetic data, to an assessment of medium to good quality real experimental STM images. The corresponding predictions are shown in figure \ref{fig: PeptideResults} d)-f). To preprocess the experimental data, the molecules of interest were first located manually. Around each selected molecule, image regions were extracted and flattened line-wise. If additional molecules were visible within the cropped region, they were removed by replacing their pixel values with the local background height, as determined by simple thresholding. 

The examples in figure \ref{fig: PeptideResults} show predictions for three different experimental \gls{STM} images of bradykinin in distinct conformations. Since the true molecular geometries and orientations are unknown, we cannot compute quantitative error metrics in this case and must instead rely on a qualitative visual evaluation of consistency between the predicted conformations and the \gls{STM} contrast.

Figure \ref{fig: PeptideResults} d) shows a molecule in the conformation labeled \textit{B} (see labels in the figure \ref{fig: PeptideResults} and Ref. \cite{rauschenbach_two-dimensional_2017}). Visually, the predicted overall molecular shape agrees very closely with the STM image. A key indicator of prediction accuracy is the placement of the two adjacent proline residues, shown in green in the right half of the molecule. Due to their relatively rigid ring geometry \cite{forster_probing_2009}, proline residues tend to protrude from the surface, giving rise to bright features in the STM image. The bright spot in the image approximately coincides with the predicted position of the second proline, supporting the correctness of the prediction. The second example, figure \ref{fig: PeptideResults} e), shows a prediction for the \textit{AB} conformation, which appears as a more compact shape. Once again, the predicted conformation aligns well with the prominent features in the \gls{STM} image. As before, the proline rings (now located toward the top of the molecule) occupy the highest positions in the image. Additionally, the aromatic rings of the phenylalanine residues correspond to distinct image features and are predicted to lie flat on the surface, as expected due to $\pi$-metal interactions \cite{mahadevi_cation-pi_2013}.
Finally, figure \ref{fig: PeptideResults} f) depicts a prediction for the conformation labeled \textit{A}. As in the previous cases, the predicted positions of the proline residues coincide with the highest-intensity features in the \gls{STM} image, and the phenylalanine-related features also show good agreement between the image and the projected conformation. The predominantly two-dimensional nature of the peptide on the surface is consistent with the strong interaction expected for adsorption on Cu(110). Across all examples, the predicted geometries mostly stick to the surface, as seen in the side views. For synthetic images, we have shown that the prediction accuracy can approach single-atom resolution. For real experimental data, however, the absence of ground-truth geometries precludes a direct quantitative assessment, and we must instead rely on identifiable structural markers in the \gls{STM} images. In the top views, key groups—particularly the phenylalanine residues, the proline rings, and the termini of the arginine side chains—are predicted with high consistency and often with near-exact placement, with only small deviations in some cases. In the side views, by contrast, we occasionally observe parts of the molecule appearing to “levitate” above or protrude into the surface. These artifacts indicate that, while the model reliably captures the overall conformation and key lateral features, further refinement steps would be beneficial to improve the vertical positioning and to move closer towards consistently accurate, atomically precise three-dimensional reconstructions for experimental \gls{STM} data.

\subsubsection*{Conformer classification}
Since we can accurately infer molecular conformations from \gls{STM} images, we can classify images of different molecules into their conformational classes \textit{A}, \textit{AB}, and \textit{B}, and into class \textit{X} for synthetic data, by classifying the predicted conformations. 

To assign conformational classes from the predicted peptide geometries, we employ a second ML workflow based on dimensionality reduction \cite{mcinnes_umap_2018} followed by a \gls{kNN} classifier (see \textit{Methods} for details).
We evaluate the classification performance on synthetic test images by comparing the predicted class with the ground-truth class used during data generation, and on real images by comparing the predictions to manually assigned labels. A detailed analysis, including confusion matrices, is provided in the \textit{Supplementary Information}. On synthetic data, we achieve a classification accuracy of \SI{95.5}{\percent} on a test set of 200 instances, underscoring the effectiveness of both the conformer prediction and the subsequent structural classification. To evaluate the performance on real data, all images present in the real dataset were manually labeled as \textit{A}, \textit{AB}, or \textit{B}. We emphasize that this labeling was performed by visual inspection and is therefore subject to human error. Additionally, this dataset contains a wide range of image qualities, complicating accurate analysis for some molecules. For the experimental images, we observe class-dependent performance: conformations \textit{A} and \textit{AB} are generally classified reliably, whereas molecules of class \textit{B} are more challenging. Their compact shape complicates accurate training-data generation and structure prediction, which in turn leads to systematic confusion with class \textit{AB}. By fine-tuning the model on an additional dataset enriched with class \textit{B} examples, we improved the overall classification accuracy on real data \SI{69}{\percent} to \SI{78}{\percent} (see \textit{Supplementary Information} for further details).

\subsection*{Glycans}
To demonstrate that our synthetic data generation and conformer prediction framework also generalizes to other classes of organic molecules, we apply the same workflow exemplarilty to a glucose hexamer, terminated with a penta-1-\-amine unit (amine linker), deposited on a Cu(100) surface using \gls{ESIBD}. The model was trained on a fully synthetic dataset consisting of 12,500 distinct molecular structures. For each molecular structure, four \gls{STM}-like images were simulated with different tip parameters, yielding 50,000 images in total. The structures which were split into training, validation, and test sets, using a 80:10:10 split.

\subsubsection*{Results on synthetic glycans}
\begin{figure*}
     \centering
    \includegraphics[width=\linewidth]{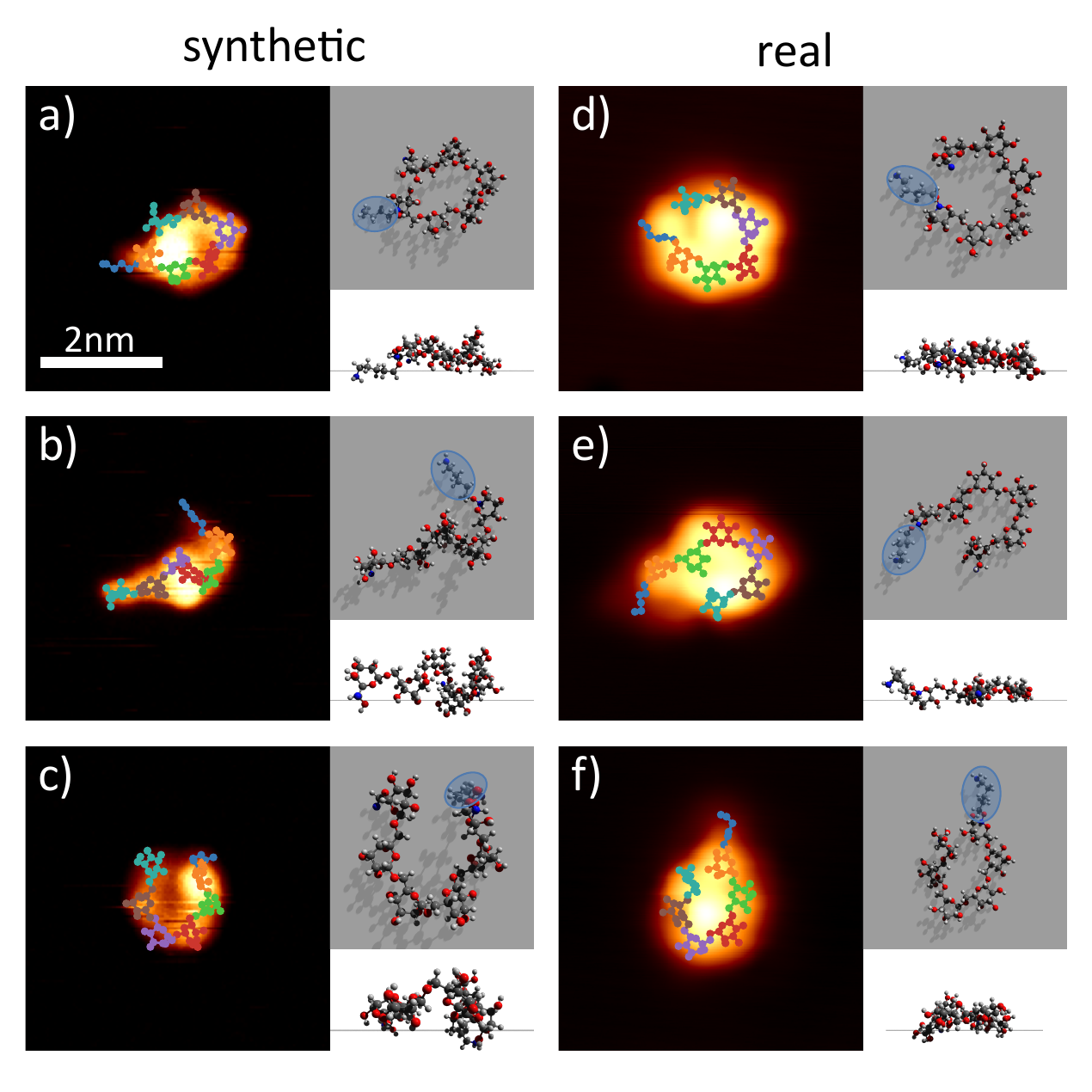}
     \caption{Results of conformer prediction model on synthetic and real images of glycans. The overlay shows the flat two-dimensional projection of the predicted atomic positions. Additionally, top and side perspectives are shown. The individual monosaccharides, and the linker are color coded and highlighted as in figure \ref{fig: Structures}. All images have a size of \SI{5}{\nano \meter} $\times$ \SI{5}{\nano \meter}}
     \label{fig: GlycanResults}
\end{figure*}

Representative results on synthetic STM-like images are shown in figure \ref{fig: GlycanResults}
a)-c). After relaxation, the synthetic glycans, consistent with experimental observations, converge to a variety of geometries, including condensed, elongated, and ring conformations. Compared to the peptides, the glycans are more flexible, observed in many different, not discretely separable conformations. Across the training samples, and especially when comparing side and top views, it is evident that glycans adopt a more three-dimensional shape than peptides, caused by their internal structure\cite{wu_imaging_2020}. This increased height variation, combined with their relatively compact lateral footprint, leads to less detailed contrast variance in both synthetic and experimental \gls{STM} data; thus, distinguishing individual features visually is more difficult. In addition, the molecular structure is dominated by identical repeating monosaccharide units, leaving the amine linker as essentially the only distinctive motif to differentiate the chain ends. The linker can typically be identified by its lower contrast in the \gls{STM} image (marked by the overlay in the structure top view). Despite these challenges, the model successfully disentangles the underlying structure in the synthetic images shown in figure \ref{fig: GlycanResults}. In a), the start of the linker is visible as a region of reduced contrast extending away from the main molecular body, while the remainder of the linker interacts with the terminal saccharide to form a more three-dimensional feature with increased height, appearing as a bright spot. Figure \ref{fig: GlycanResults} b) shows a more elongated conformation, in which the linker-terminated end of the chain can be distinguished from the opposite end by its different contrast. The third example, figure \ref{fig: GlycanResults} (c), illustrates a ring-like conformation, where the topography forms a circular feature at the location where the linker interacts with the chain end, again giving rise to a local height maximum. Thus, even though the amine linker is the only asymmetric structural element and the pronounced three-dimensionality degrades image clarity and complicates structural interpretation, the conformer prediction model nonetheless provides accurate reconstructions of the overall mole\-cular shape across the examples shown. Extending the analysis, as for peptides, to the level of individual atoms, we still obtain good localization performance: the median atomic position error across the synthetic glycan test set is \SI{3.5}{\angstrom}, \SI{20}{\percent} have an error below \SI{2.5}{\angstrom}, \SI{80}{\percent} an error below \SI{6.3}{\angstrom}.

\subsubsection*{Results on real glycans}

Applications to real \gls{STM} images of glycans are shown in figure \ref{fig: GlycanResults} d)-f). In the first example d), the glycan is adsorbed in a relatively flat conformation, yielding comparatively detailed intermolecular structural details and allowing individual monosaccharide units to be distinguished. Their positions, as well as the location of the amine linker, agree well with the predicted geometry; only the terminal monosaccharide may, in reality, be curled slightly more towards the molecular center than in the prediction. In the more compact second and third examples (e) and f)), not all monosaccharide units are clearly discernible by eye. Nevertheless, the prediction model infers plausible conformations that reproduce the key image features and overall molecular shapes, thereby delivering accurate global geometries. This demonstrates that, even for complex three-dimensional molecules that consequently have less intermolecular structurally resolved detail in \gls{STM} imaging, the model can still predict realistic conformations from real \gls{STM} images. 

\section*{Discussion}

We have presented a complete framework for fully automated estimation of macro-molecular conformations from \gls{STM} images of individual peptide, glycan, and glycopeptide molecules. The framework comprises two main components: (i) a fast, high-quality, and highly diverse training-data generation pipeline for arbitrary glycans and peptides, and (ii) a \glsentrylong{CNN}–based conformer prediction model. 

The training data are generated using a custom, highly flexible script that assembles peptides and glycans in a modular, piecewise fashion and relaxes the resulting structures under the inclusion of an approximate surface potential. This strategy enables the creation of a broad geometric variety of molecules for synthetic data sets, as demonstrated for both peptide and glycan systems. When reference conformations are available, they can be exploited to additionally bias the sampling toward experimentally relevant structures, as we did for the peptide conformations. By combining state-of-the-art \gls{ML}-based electronic density prediction\cite{del_rio_deep_2023} with a custom \gls{STM} simulation, we rapidly generate diverse training sets for both example molecules, with a total computation time of less than 10 seconds per image on a single GPU.

For structure inference, we proposed an \gls{ML} framework that uses a ResNet-based encoder\cite{he_deep_2015} coupled with a regression head to predict atomic pairwise distances directly from \gls{STM} images. These distances can then be used to reconstruct three-dimensional atomic positions. Training and evaluation on synthetic datasets show that the resulting models can accurately recover the overall conformations of large, three-dimensional mole\-cules. When applied to real \gls{STM} images of poly\-peptides and glycans, the models yield convincing structures in which key image features are consistently mapped to the corresponding molecular subunits.

From a machine-learning perspective, realistic, variable, and fast data generation remains a critical bottleneck for many \gls{AI} applications in \gls{STM}, \gls{AFM}, and \newline\gls{STEM}\cite{li_machine_2021, seifert_chirality_2024, masud_machine_2024, botifoll_machine_2022}. The data generation scheme proposed here provides a promising solution for \gls{STM} and is sufficiently flexible to be extended to other imaging techniques, such as \gls{AFM}, and to different tasks, including object detection and semantic segmentation. As such, it has the potential to accelerate many common data-analysis workflows, support a variety of \gls{ML} tasks, and substantially reduce the need for manual labeling and interpretation. Furthermore, our results show that a standard \gls{CNN} architecture, when paired with an efficient encoding and a suitable regression head, can produce highly accurate structural predictions. Owing to the encoder–decoder design, the \gls{CNN} backbone can, in principle, be replaced with more advanced computer-vision mo\-dels if required by task complexity. 

From a physics standpoint, the method bridges the gap between inaccurate, labor-in\-ten\-si\-ve, hu\-man-dependent manual structural analysis and fully quantum-mechanical approaches such as\newline\gls{DFT} and \gls{MD}, which are accurate but extremely time- and resource-in\-ten\-si\-ve for such macro-molecular complexes. \gls{DFT} and \gls{MD} workflows also rely on manually constructed initial guesses of the molecular structure, and the quality of these guesses strongly affects the outcome\cite{kudin_black-box_2002}. Our framework can automate and improve the generation of such initial structures, opening the door to more fully automated structure-determination pipelines. 

When applied to experimental images, our method has shown to robustly transfer from  purely synthetic training data to practical applications to experimental images and surpasses existing approaches\cite{kurki_automated_2024, carracedo-cosme_molecular_2024}.
On synthetic benchmarks, we showed that the model can evaluate proposed conformations down to the level of individual atomic positions. In the present implementation, several simplifications and approximations were necessary to balance physical realism with computational efficiency. 

In its current state, our framework requires a priori knowledge about the molecular structure to disentangle its conformation. The possibility of generating synthetic images of various peptides, glycans, and glycopeptides however opens the path to apply different \gls{ML} tasks, such as classifying different species of molecules, solely restricted by the atomic components learned by the \gls{ML}-\gls{DFT} framework\cite{del_rio_deep_2023}.
Looking at larger molecules, the training data generation scales linearly with the number of atoms in the molecule. Since we encode our molecules as a distance matrix, the parameters in the output layer of our network scale quadratically with the number of features to locate in the image (here individual atoms, but can also be chosen as single monosaccharides or amino acids). While this will be a limiting factor for larger molecules, it is still a significant improvement over conventional \gls{DFT}, which scales with the third power of atoms in the system \cite{del_rio_deep_2023}.
Furthermore, enhancing the underlying physical modeling, particularly by incorporating more accurate molecule–surface interactions, should bring the synthetic training data into closer agreement with experiments and further improve prediction quality on real images, moving toward a fully automated, atomically precise structure solver.

In conclusion the proposed framework accurately predicts molecular conformations from both synthetic and experimental STM images of biologically relevant three-dimensional macro-mo\-le\-cu\-les and thus goes far beyond all approaches available for small individual molecules. Its de\-mon\-stra\-ted flexibility for diverse glycans and peptides represents a substantial simplification for structural studies of biomolecules and will facilitate single-molecule imaging approaches that are crucial for advancing our understanding of biological systems. 

\section*{Methods}

\subsection*{Synthetic Conformer generation}
The conformer generation is implemented for arbitrary peptides, glycans, and glycopeptides. By parsing the provided structure, the molecule is assembled piece by piece in iterative steps. First, the geometry for a new block, either an amino acid or a glycan, is created using standard interatomic distances and positions. After initializing a second block, it is stitched to the existing chain, following three randomly sampled angles. Modifying the probability distribution of alignment angles yields different conformational classes. To relax the structure internally and simulate surface adsorption, the structure is first modified using gradient descent on a Lennard-Jones potential modeling the surface, followed by relaxation using the RDkit implementation of the \glsfirst{UFF}. These steps are repeated while increasing the number of gradient steps along the surface potential using a sine function and decreasing the \gls{UFF} relaxation steps. To this semi-relaxed structure, the next block is attached. This process is repeated until the molecular assembly is complete.
The conformer generation process is fully automated. However, numerical instabilities during optimization can occasionally result in nonphysical structures, most notably steric clashes. Such conformations are detected and removed by an automated post-processing filter according to their final \gls{UFF} energy, their surface energy, and minimum and maximum a\-to\-mic bond distances.

\subsection*{Electron Density Prediction}
To simulate accurate \gls{STM} images, the electronic density for a given molecular geometry needs to be determined. As a workaround for computationally very demanding \gls{DFT} calculations, \textit{del Rio et al.}\cite{del_rio_deep_2023} have proposed and trained a \gls{ML} based framework that emulates \gls{DFT} calculations on over 100,000 molecules, polymer crystals, and polymer chains composed of nitrogen, carbon, oxygen, and hydrogen atoms. 
The implementation published by \textit{del Rio et al.} has been modified to match the computation speed demands for streamlined data generation: By offloading the large matrix calculations encoding the molecular geometry to the GPU, we were able to achieve \gls{DFT} computation times below $\SI{10}{\second}$ for our exemplary structures, at resolutions below $100 \times 100 \times 100 \text{px}$.

\subsection*{STM-like image simulation}
For increased variation in the training data, we sample four \gls{STM} setups per structure, randomly selecting tip and scanning parameters such as tip shape and setpoint. The tip is modeled as a stack of two-dimensional Gaussian functions with varying radii, tilt, and eccentricity. We calculate a discrete three-dimensional convolution of the tip and the electronic density, interpreting the overlap at each point in space as equivalent to a tunneling current. At the substrate plane defined during geometry generation, we model the surface charge density as a logistic function. Using a PID controller, we scan \gls{STM}-like images at a randomly sampled setpoint, simulating constant current \gls{STM} operation. Synthetic example images can be found in the \textit{Supplementary Information \ref{SI_Data}}.

\subsection*{STM measurements}
The experimental \gls{STM} images discussed have been gathered using two different microscopes. Peptide measurements have been performed by \textit{Rauschenbach et al.} Experimental details can be found in the literature \cite{rauschenbach_two-dimensional_2017}.

Glycans were soft landed on an atomically flat Cu(100) (obtained by repeated cycles of Ar+ sputtering and annealing at \SI{790}{\kelvin}) in \gls{UHV} using a home-built \gls{ESIBD} apparatus at the Max Planck Institute for Solid State Research, Stuttgart, Germany. The Cu(100) surface was held at \SI{80}{\kelvin} during the deposition process. The base pressure during the ion beam deposition at the surface was \SI{<5e-10}{\milli\bar}. The prepared surface with the glycans were then transferred to a self-built low-temperature (\SI{4}{\kelvin}) \gls{STM} setup in \gls{UHV}. The glycans were imaged under constant-current imaging conditions using a typical tunnelling current setpoint of \SI{0.1}{\pico \ampere} to \SI{5}{\pico \ampere} with a tunneling bias of \SI{0.3}{\volt}. 

\subsection*{Machine Learning Setup}
For conformer predictions, hydrogen atoms are neglected, thereby saving computational re\-sour\-ces. The peptide molecules are composed of 76 non-hydrogen atoms, and the glycan molecules are composed of 73 non-hydrogen atoms, containing carbon, nitrogen, and oxygen. The molecular geometries are encoded as distance matrices; hence, absolute position and translation information are lost. To preserve this, we incorporate an anchor structure of 13 coordinates at fixed points in space at 
$
    \vec{x}= x_i \cdot \vec{e}_{\{x,y,z\}}, \hspace{3pt} x_i \in 
    \{ 0,\pm\SI{5}{\angstrom},\pm\SI{10}{\angstrom}\}
$. The distance matrices, therefore, contain $89^2$ and $86^2$ entries, respectively. However, since this is a symmetric matrix with a zero diagonal, the total number of parameters reduces to $3916$ entries for peptides and $3655$ for the example glycan.
The \gls{ML} network follows an encoder-decoder scheme. We selected the 50 layer \gls{CNN} ResNet50\cite{he_deep_2015} for our encoder, pretrained on the ImageNet dataset\cite{deng_imagenet_2009}. We drop the last fully connected layer to reduce the bottleneck and add a custom decoder model consisting of two fully connected layers to expand the dimension to $4096$ parameters, as well as a final fully connected layer with sigmoid activation and the number of output parameters required by the conformer for prediction. 

\subsubsection*{Data Augmentation}
Since the produced synthetic \gls{STM} images are free of defects, further data augmentation is required. The images are shifted and scaled in the xy-direction, compensating for calibration mismatches in the experimental application. The image is then randomly flipped, rotated, and scaled to a fixed resolution of $128\times \SI{128}{\pixel}$.
We improve the realism of the images by adding \gls{STM}-typical scanning noise, shot noise, and perlin noise\cite{perlin_improving_2002} for the background. The peptide dataset, furthermore, contains artificial oscillations that are occasionally seen in the background of some real images. Contrast is enhanced in the high values using an exponential function, and the image is normalized using min-max normalization.

\subsubsection*{Training Setup}
The peptide detection model has been trained for 50 epochs on 4,800 different structures, hence 14,200 different \gls{STM}-like images. The model for glycans was trained for 20 epochs on 10,000 different structures, thus 40,000 \gls{STM} images. The training hyperparameters are identical for both tasks. We use a batch size of 40 and the \textit{AdamW} optimizer\cite{loshchilov_decoupled_2019}; the learning rate is adapted using OneCylce\cite{smith_super-convergence_2018} with a peak learning rate of $0.0003$. The loss function consists of three contributions: First, a structural loss as the mean squared error between the predicted and target training distance matrices, aimed at predicting the correct conformation and position. Second, a surface loss is calculated as the Lennard-Jones potential of the structure with respect to the surface, enforcing flat and level molecules, as is physically expected. Lastly, we include a steric loss, penalizing very close atomic distances with a quadratic potential. The weight of each loss is adapted dynamically, following the ideas of GradNorm\cite{chen_gradnorm_2018} and Dynamic Weight Averaging\cite{liu_end--end_2019}, initially setting almost complete attention to the structure loss.

\subsubsection*{Structure Reconstruction}
Given the predicted distance matrix, we need to be able to reconstruct the conformation and its orientation. We use the scikit-learn SMACOF implementation for metric \gls{MDS}\cite{pedregosa_scikit-learn_2018, kruskal_multidimensional_1964, borg_modern_1997, mair_more_2022}. Given the fixed anchor structure and its reconstruction, we can align the mo\-le\-cu\-le within the image using the \textit{Kabsch} al\-go\-ri\-thm\cite{kabsch_solution_1976}.
The reconstruction based on \gls{MDS} also provides a step of unsupervised validation: \textit{Kruskal} calculates the deviation between reconstructed and provided distances as the \gls{STRESS} and quantifies the reconstruction accuracy according to this value, as further discussed in the \textit{Supplementary Information \ref{SI_STRESS}}

\subsubsection*{Conformer Classification}
Exemplarily for the peptide structure, we demonstrate that the predicted conformation can further be used to assign the conformations into different classes. After scaling to a zero mean and standard variance, we can create a lower-dimensional embedding of the distance vectors of our training samples. We chose \gls{UMAP} \cite{mcinnes_umap_2018} as our dimensionality reduction technique, embedding distance matrices into a three-di\-men\-sio\-nal latent space. As further discussed in the \textit{Supplementary Information \ref{SI_Latent}}, we observe distinct, separate clusters corresponding to different conformational classes. These could be separated automatically using unsupervised learning, thereby creating automatic classifications. Since we know the conformer class of the training samples, we can use their embedding to classify predictions on new data. Given a new synthetic or real predicted structure, we can scale and embed the distance matrix into the same latent space as done for our training samples since \gls{UMAP} supports out-of-sample embedding. The new latent vector can be compared to the training samples to infer a class. 
For this, we use \gls{kNN} classification with the BallTree algorithm \cite{omohundro_five_1989} as implemented by scikit-learn\cite{pedregosa_scikit-learn_2018}, using $k=11$.

\bibliographystyle{unsrtnat}
\bibliography{references2}

\newpage
\clearpage
\onecolumn
\section{Supplementary Information}
\glsresetall

\subsection{Variety of simulated \gls{STM} images}
\label{SI_Data}

The \gls{STM} simulation framework allows us to generate a vast variety of different conformations given a single peptide, glycan, or glycopeptide structure and to randomly select settings to generate a number of different \gls{STM} images per structure.

We show this for three examples: a peptide sequence of Angiotensin II in figure \ref{fig: SI_STM_peptide}, a glycopeptide structure in figure \ref{fig: SI_STM_glycopeptide}, and a oligosaccharide in figure \ref{fig: SI_STM_glycan}

\newcommand{\imgw}{0.29}
\begin{figure}[!htbp]
    \centering 
    \begin{tabular}{@{}c@{}c@{}c@{}} 
        \includegraphics[width=\imgw\linewidth]{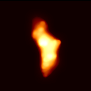} &
        \includegraphics[width=\imgw\linewidth]{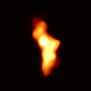} &
        \includegraphics[width=\imgw\linewidth]{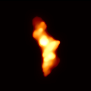} \\
        \includegraphics[width=\imgw\linewidth]{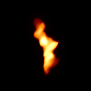} &
        \includegraphics[width=\imgw\linewidth]{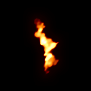} &
        \includegraphics[width=\imgw\linewidth]{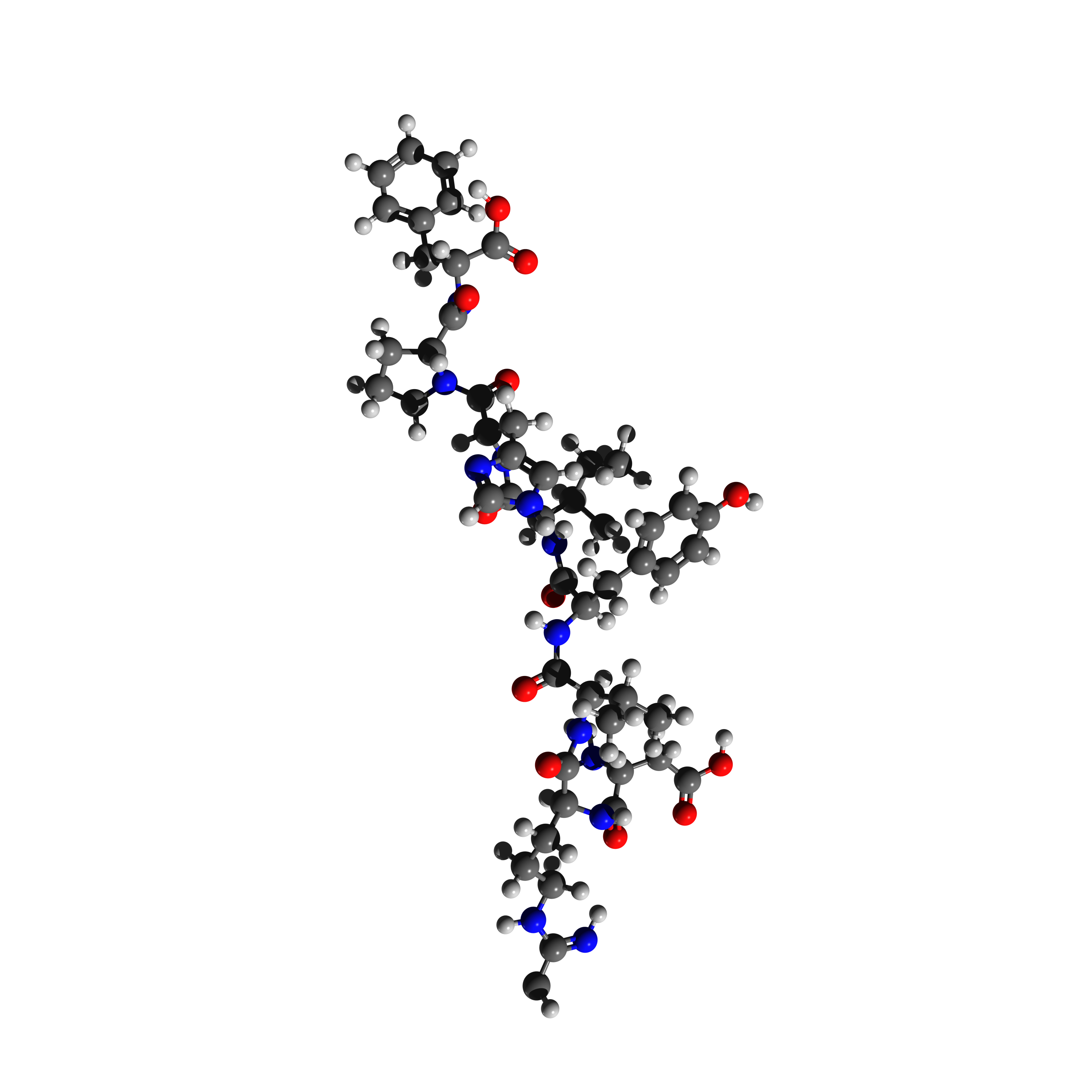}
    \end{tabular}
    \caption{Example \gls{STM} images of the peptide Angiotensin II with sequence ASP-ARG-VAL-TYR-ILE-HIS-PRO-PHE. Figures show different synthetic \gls{STM}-like images of the same structure, resulting from the identical predicted electronic density, as well as a three-dimensional top-down view of the generated conformation\cite{hanwell_avogadro_2012}.}
    \label{fig: SI_STM_peptide}
\end{figure}

\begin{figure}[!htbp]
    \centering 
    \begin{tabular}{@{}c@{}c@{}c@{}} 
    
        \includegraphics[width=\imgw\linewidth]{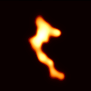} &    
        \includegraphics[width=\imgw\linewidth]{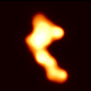} &
        \includegraphics[width=\imgw\linewidth]{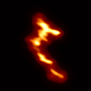} \\
        \includegraphics[width=\imgw\linewidth]{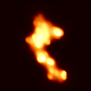} &
        \includegraphics[width=\imgw\linewidth]{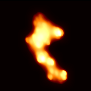} &
        \includegraphics[width=\imgw\linewidth]{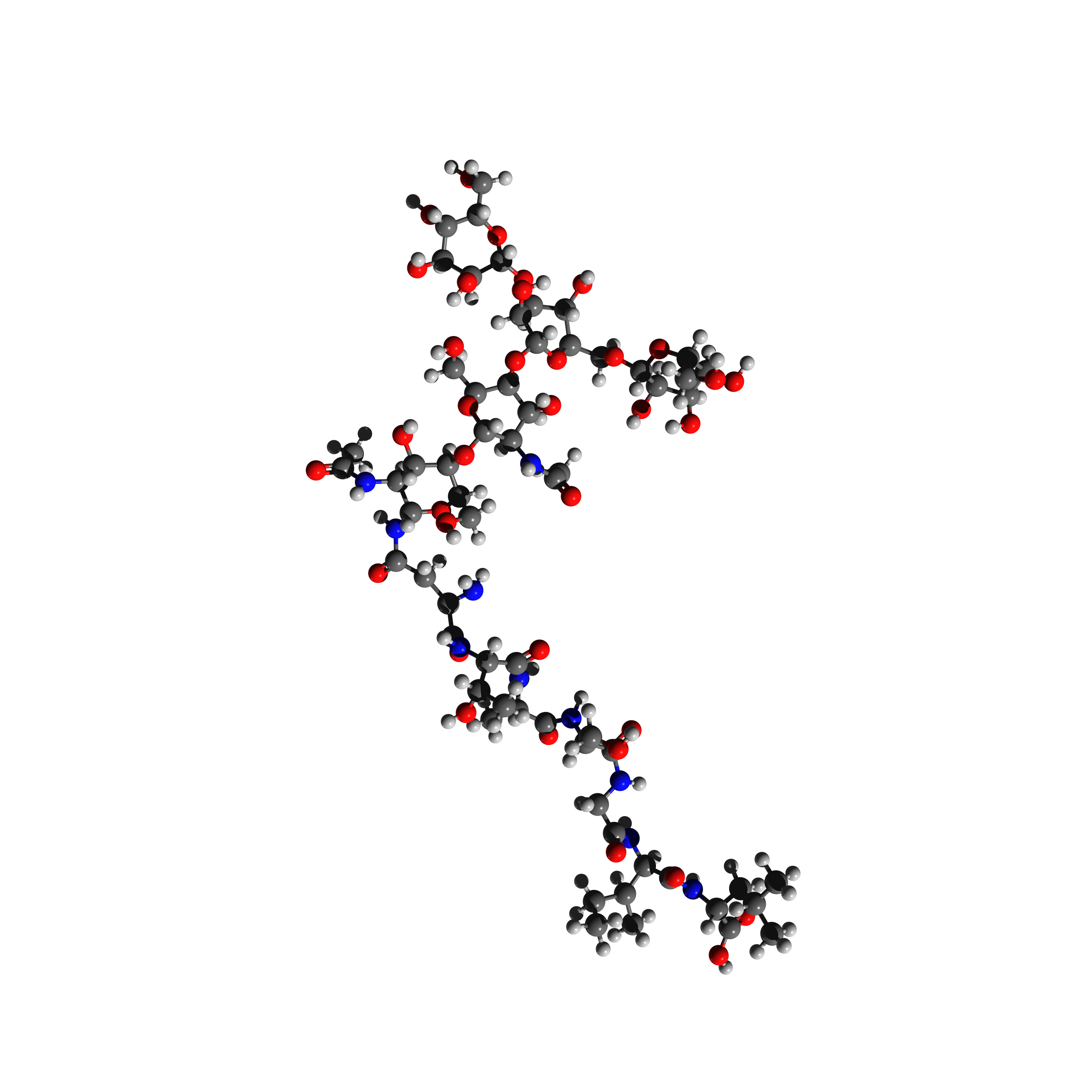}
    \end{tabular}
    \caption{Example \gls{STM} images of a glycopeptide. The peptide sequence is ASN-THR-ALA-SER-GLY-ILE-LEU. Connected to the asparagine is a  N-glycan core oligosaccharide consisting of three mannose residues in a branched form, attached to a mannose–N-acetylglucosamine–N-acetylglucosamine chitobiose core.
    Figures show different synthetic \gls{STM}-like  images of the same structure, resulting from the identical predicted electronic density, as well as a three-dimensional top-down view of the generated conformation\cite{hanwell_avogadro_2012}.}
    \label{fig: SI_STM_glycopeptide}
\end{figure}

\begin{figure}[!htbp]
    \centering 
    \begin{tabular}{@{}c@{}c@{}c@{}} 
    
        \includegraphics[width=\imgw\linewidth]{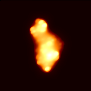} &
        \includegraphics[width=\imgw\linewidth]{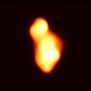} &
        \includegraphics[width=\imgw\linewidth]{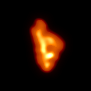} \\
        \includegraphics[width=\imgw\linewidth]{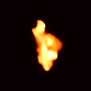} &
        \includegraphics[width=\imgw\linewidth]{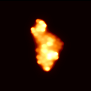} &
        \includegraphics[width=\imgw\linewidth]{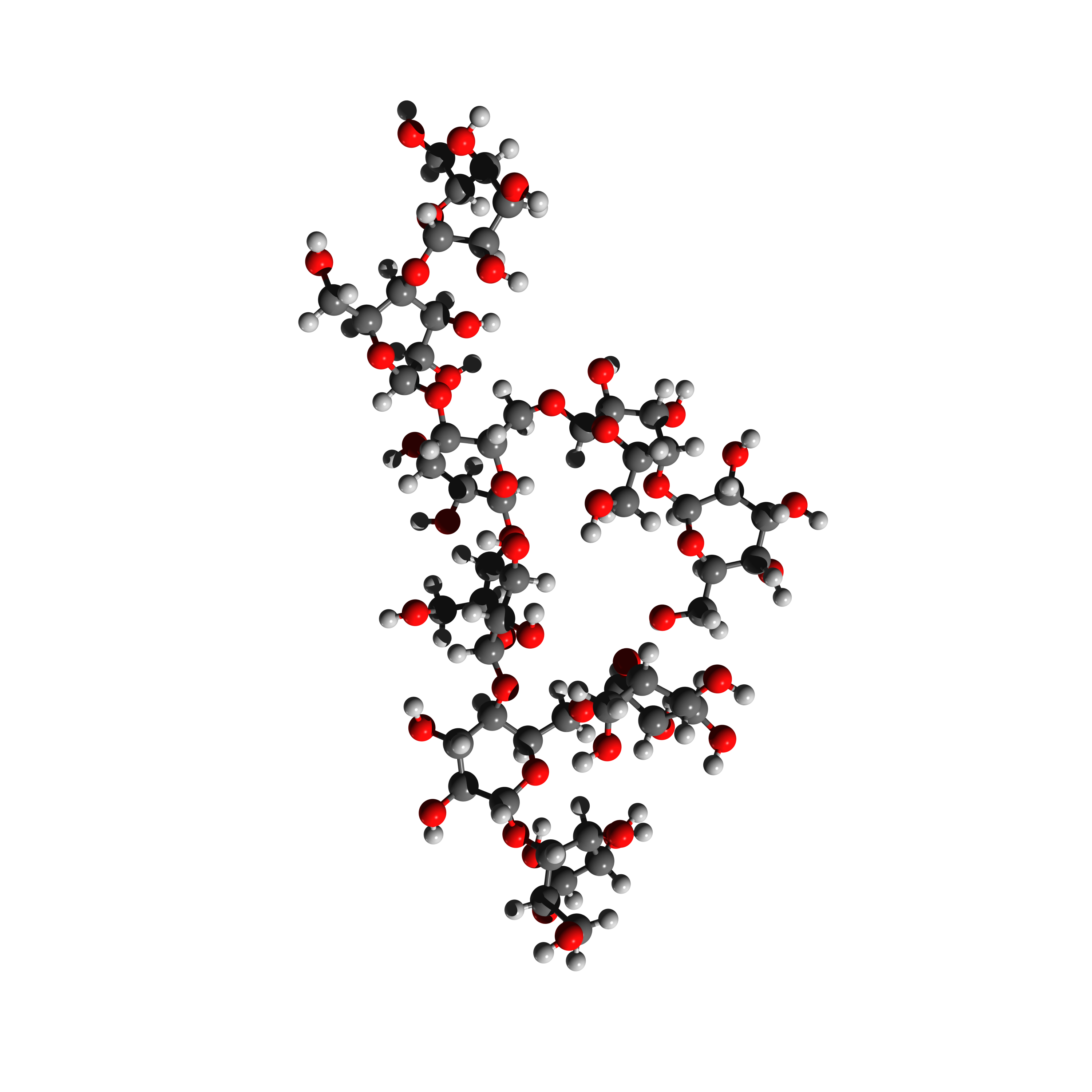}
    \end{tabular}
    \caption{Example \gls{STM} images of a glycan. The molecule consists of nine glucose units, assembled in a main chain with six units and two branches.
    Figures show different synthetic \gls{STM}-like  images of the same structure, resulting from the identical predicted electronic density, as well as a three-dimensional top-down view of the generated conformation\cite{hanwell_avogadro_2012}.}
    \label{fig: SI_STM_glycan}
\end{figure}

\clearpage
\subsection{Statistics on training data}
\label{SI_TrainDataStat}

\begin{figure}[!htbp]
    \centering 
    \includegraphics[width=\textwidth]{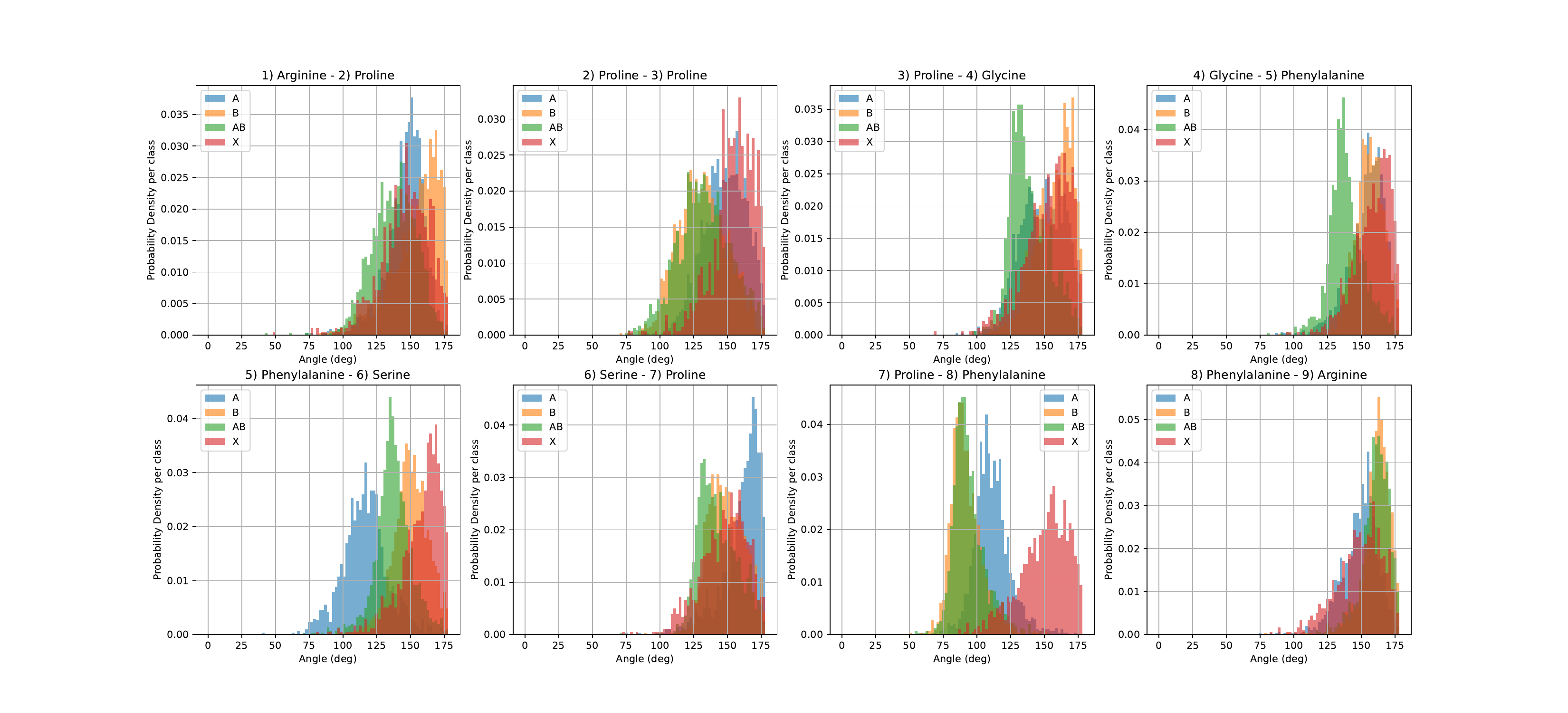}
    \caption{Distribution of final angles between adjacent amino acids in the generated training structures, depending on the assigned conformer class. The angle between two amino acids is determined by the angle between the vectors describing the backbone of each amino acid, given by the vector between the nitrogen position, and the last carbon atom.}
    \label{fig: SI_Stat_Pep_angle}
\end{figure}

\begin{figure}[!htbp]
    \centering 
    \includegraphics[width=0.7\textwidth]{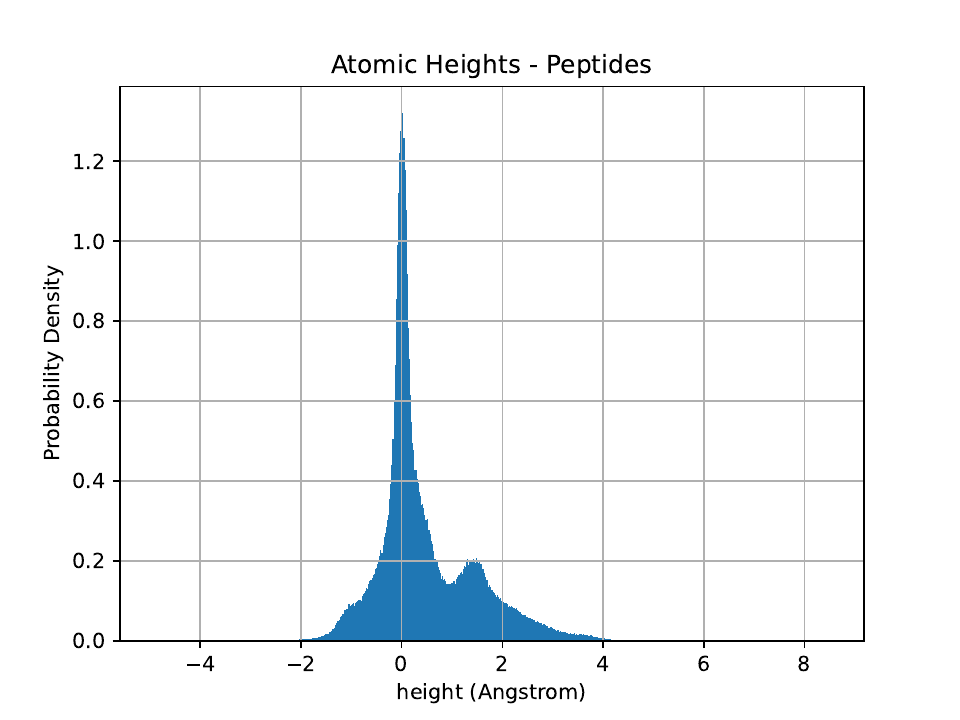}
    \caption{Distribution of atomic heights in the generated peptide data. The height $z=\SI{0}{\angstrom}$ denotes the energetic minimum of the energy function. 90 \% of atoms are located in the interval between a height of \SI{-0.8}{\angstrom} and \SI{+2.4}{\angstrom}. }
    \label{fig: SI_Stat_Pep_height}
\end{figure}

\begin{figure}[!htbp]
    \centering 
    \includegraphics[width=0.8\textwidth]{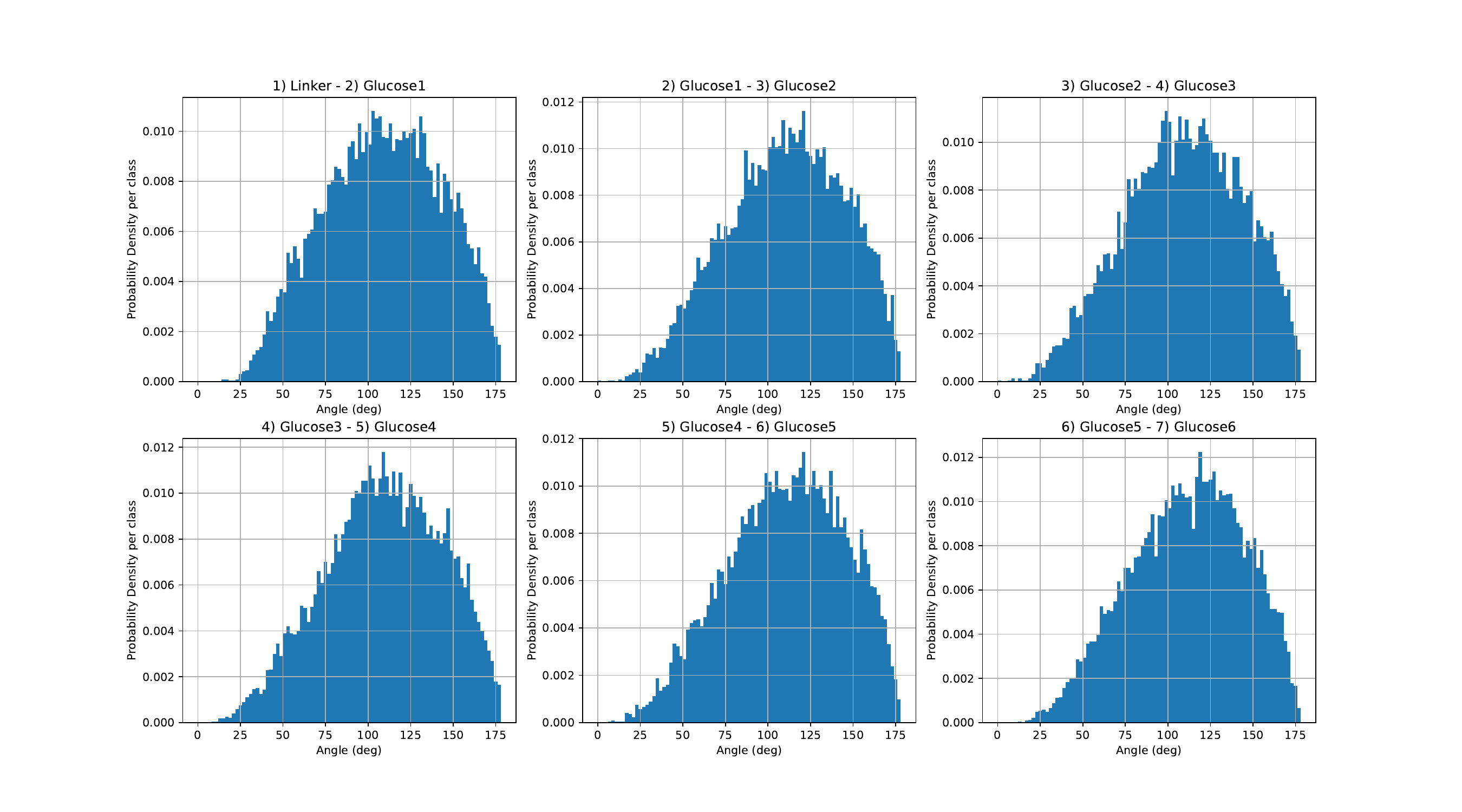}
    \caption{Distribution of final angles between adjacent amino acids in the generated training structures. The angle between two monosaccharides is determined by the angle between the vectors describing the last carbon and first oxygen atoms of each monosaccharide}
    \label{fig: SI_Stat_Gly_angle}
\end{figure}

\begin{figure}[!htbp]
    \centering 
    \includegraphics[width=0.7\textwidth]{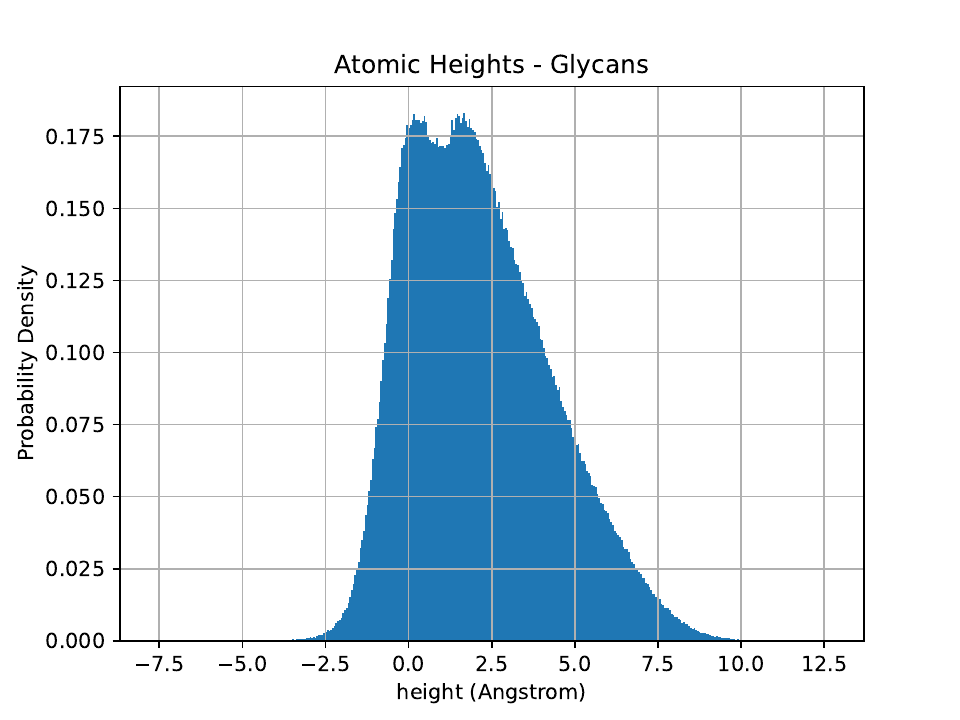}
    \caption{Distribution of atomic heights in the generated glycan data. The height $z=\SI{0}{\angstrom}$ denotes the energetic minimum of the energy function. 90 \% of atoms are located in the interval between a height of \SI{-0.8}{\angstrom} and \SI{+6.0}{\angstrom}. } 
    \label{fig: SI_Stat_Gly_height}
\end{figure}


\subsection{Latent space for peptide classification}
\label{SI_Latent}

The peptide dataset has been refined by changing the bond angle distributions to better match the observed results. This distribution is split into multiple modes: three preferred alignments are constructed to match the observed classes \textit{A}, \textit{AB}, and \textit{B}. Additionally, we denote a third class, \textit{X}, to geometries where all angles are chosen from a uniform distribution .

\begin{figure}[!htbp]
    \centering
    \begin{subfigure}[b]{0.45\textwidth}
        \centering
    \includegraphics[width=\textwidth]{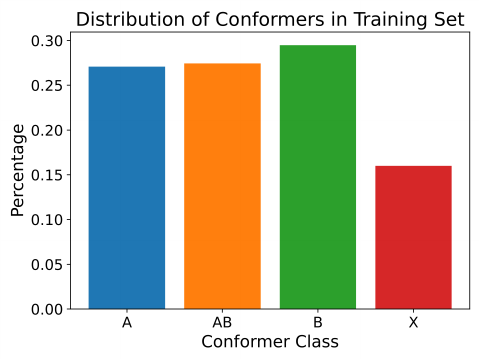}
    \caption{}
        \label{fig:SI_ConfDistro}
    \end{subfigure}
    \hfill
    \begin{subfigure}[b]{0.45\textwidth}
        \centering
    \includegraphics[width=\textwidth]{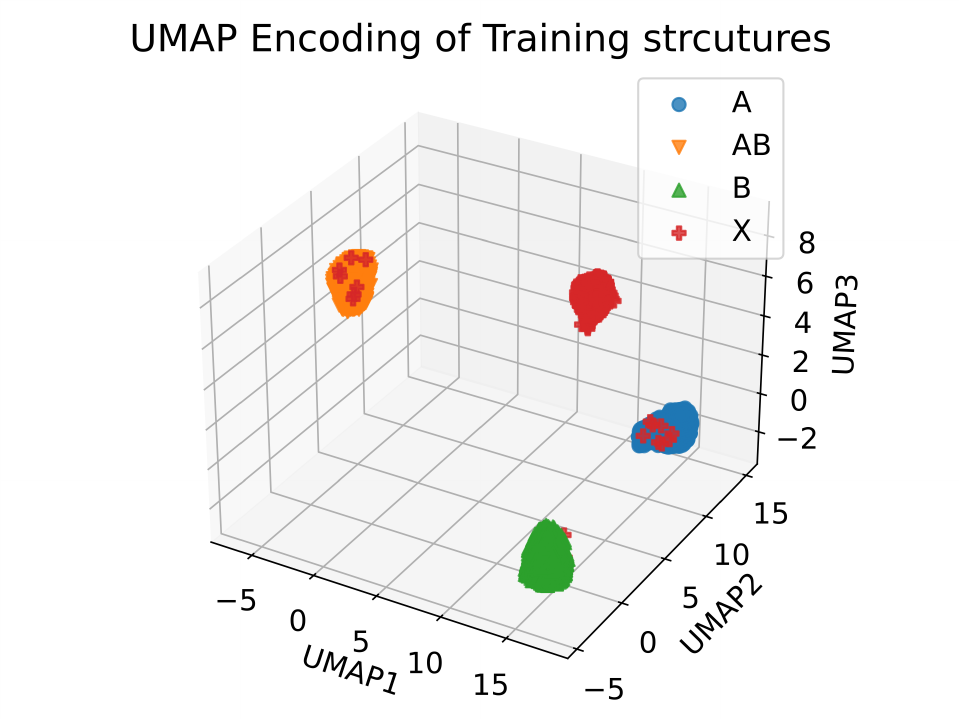}
        \label{fig:SI_UmapEncoding}
        \caption{}
    \end{subfigure}
    \caption{a) Distribution of different conformer classes in the peptide training dataset. b) Visualized UMAP encoding of the structure distance matrices in the training dataset, color coded according to the conformer class.}
    \label{fig:SI_Encoding}
\end{figure}

Figure \ref{fig:SI_Encoding} a) shows the distribution of the different conformer classes in the training dataset. During the training data generation, the class for each new molecule is sampled randomly and equally from the three preset classes, as well as from the free form generation scheme labeled as \textit{X}. Deviation from equal distribution can occur due to different generation times and generation success probabilities. This is especially the case for class X: Since the bond angles are selected completely at random, there is a higher chance that steric collisions occur. The structural relaxation in this case needs more iterations and, thus, more time to succeed. If the collision is not easily resolvable, the angles will be resampled, leading to an overall longer structure generation time for class X compared to the others, resulting in fewer examples of this class.

Figure \ref{fig:SI_Encoding} b) shows the three-dimensional latent space of the learned UMAP encoding of all training samples, encoded by their conformation class. It can be seen that the different classes show well-distinct clusters, thus being robustly separable by the UMAP encoder. In some cases, instances of class \textit{X} converge towards the pre-determined classes, resulting in isolated instances of class X in the clusters of all other classes.  For this reason, we use a high k-value of 11 for the \textit{kNN} classifier, aiming to overlook those instances.

Note that the UMAP latent space is highly nonlinear. As a result, the conformers of class X appear very close, compact, and not spread across the entire probability space. This is, however, only the case in this embedding, as the conformations are actually very diverse and become apparent through instances of class \textit{X} located in all other clusters.

\subsection{Validation via \gls{MDS}}
\label{SI_STRESS}

\gls{MDS} also provides a step of unsupervised validation: \textit{Kruskal} calculates the deviation between reconstructed and provided distances as the \gls{STRESS}, and quantifies the reconstruction accuracy according to this value, labeling \gls{STRESS} values below $\SI{5}{\percent}$ as a \textit{good} fit \cite{kruskal_multidimensional_1964}. Calculating the distance matrix for the input structures and reconstructing the geometry can be done very accurately, achieving \gls{STRESS} values below $\SI{0.5}{\percent}$. In the predicted distance matrices, the distances between pairwise atoms are all predicted independently of one another. As a result, the estimated distance matrix, however, will inevitably contain internal contradictions. Hence, even with a perfect algorithm, a reconstruction satisfying the complete distance matrix is impossible, resulting in increased \gls{STRESS}. As the structure prediction improves during training, internal contradictions are reduced, thus lowering the \gls{STRESS}. For the synthetic test dataset of peptides, we achieve an average \gls{STRESS} of $\SI{4}{\percent}$, thus indicating good quality reconstructions according to \textit{Kruskal}.

\subsection{Peptide classification results}
\label{SI_Classification}

\begin{figure}[!htbp]
    \centering
    \begin{subfigure}[b]{0.45\textwidth}
        \centering
    \includegraphics[width=\textwidth]{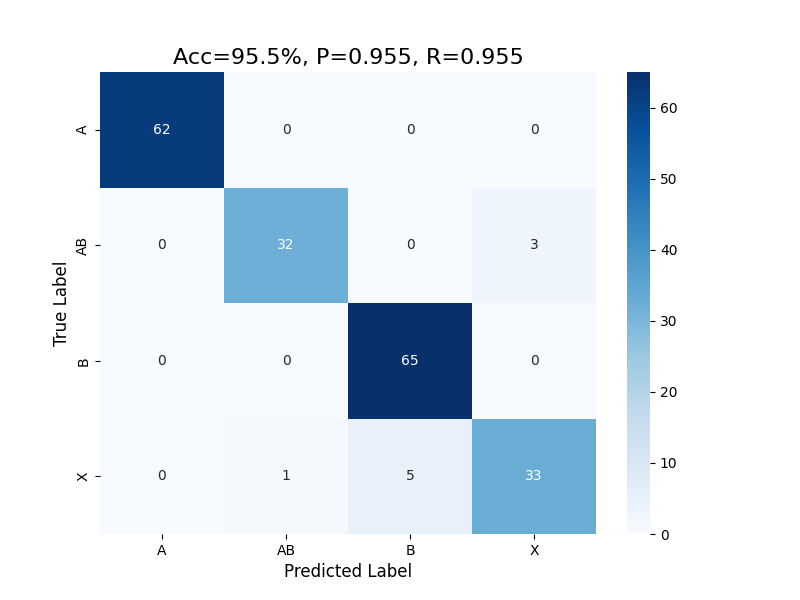}
    \caption{Synthetic Data}
        \label{fig:ClassSyn}
    \end{subfigure}
    \hfill
    \begin{subfigure}[b]{0.45\textwidth}
        \centering
    \includegraphics[width=\textwidth]{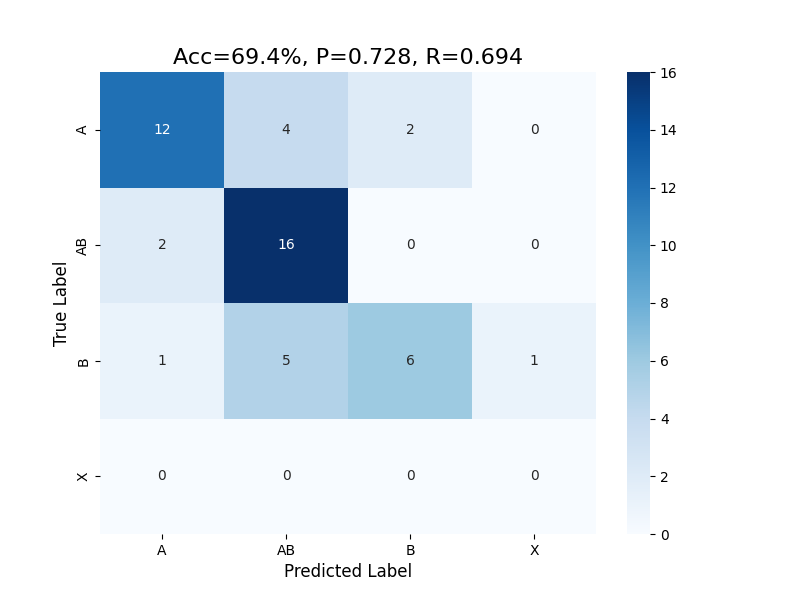}
        \label{fig:ClassReal}
        \caption{Real Data}
    \end{subfigure}
    \caption{a) Confusion matrix for conformer classification on synthetic images b) Confusion matrix for conformer classification on real images.}
    \label{fig:Classification}
\end{figure}

We constructed confusion matrices for both cases, visualized in figure \ref{fig:Classification}.

The confusion matrix for synthetic images (\ref{fig:Classification} a)) highlights the success of our method on synthetic images: Within the total test set of 201, only 9 instances were classified incorrectly, leading to an accuracy of $\SI{95.5}{\percent}$.

\subsection{Finetuning Detection Model}
\label{SI_Finetuning}

\begin{figure}[!htbp]
    \centering
    \includegraphics[width=\textwidth]{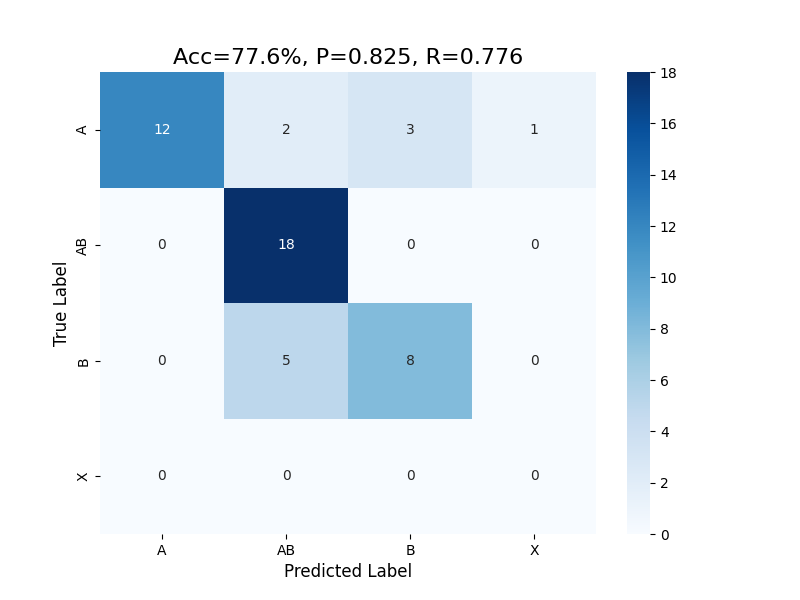}
    \caption{Confusion matrix for real images after Finetuning}
    \label{fig:SI_FTClass}
\end{figure}

When applying the peptide conformer prediction model, which was trained on synthetic data, to real images and conducting the conformer classification, we observed an imbalance in prediction accuracy.
Although classes \textit{A}, \textit{AB}, and \textit{B} are represented equally in the training dataset (see figure \ref{fig:SI_Encoding} a)), the detection accuracy for class \textit{B} was significantly inferior to that of the others, potentially due to a stronger folding; thus, the images lack detail in the folded part.
As an attempt to improve the detection accuracy for class \textit{B}, we finetuned the conformer prediction model.
We generated a second dataset containing 3,500 training images, but this time with an imposed class imbalance of $\SI{80}{\percent}$ favoring class \textit{B}. We continue to train the ResNet-based model on this dataset for 50 epochs. As for hyperparameters, we reduce the learning rate by a factor of 10. 
With this training, we were able to increase the overall classification accuracy from $\SI{69}{\percent}$ to $\SI{78}{\percent}$, enhancing the accuracy of classes \textit{AB} and \textit{B} allowing the model to better differentiate between these classes. The confusion matrix is shown in figure \ref{fig:SI_FTClass}.

\clearpage
\end{document}